\documentclass[letterpaper,12pt,amsfonts,amssymb]{article}

\usepackage{graphics,graphicx,psfrag,color,float, hyperref}
\usepackage{epsfig}
\usepackage{amsmath}
\usepackage{amssymb}

\usepackage{amsmath}
\usepackage{amssymb}
\usepackage{amsfonts}
\usepackage{amsthm, hyperref}
\usepackage {times}
\usepackage{bbm}
\usepackage{braket}

\newcommand{\beq}{\begin{equation}}
\newcommand{\enq}{\end{equation}}
\newcommand{\bel}{\begin{lemma}}
\newcommand{\enl}{\end{lemma}}
\newcommand{\bet}{\begin{theorem}}
\newcommand{\ent}{\end{theorem}}
\newcommand{\ten}{\textnormal}

\newcommand{\tr}{\mathrm{Tr}}
\newcommand{\myexp}{{\mathrm{e}}}

\newtheorem{theorem}{Theorem}
\newtheorem{lemma}{Lemma}

\title{On some special cases of the Entropy Photon-Number Inequality}
\author{Smarajit~Das, Naresh~Sharma,
and~Siddharth~Muthukrishnan\thanks{S. Das, N. Sharma and S. Muthukrishnan are with the School of Technology and
Computer Science,
Tata Institute of Fundamental Research, Mumbai 400 005, India. email:\{smarajit, nsharma\}@tifr.res.in, smkrish@tcs.tifr.res.in.}}

\date{\today}

\begin {document}

\maketitle

\abstract{
We show that the Entropy Photon-Number Inequality (EPnI) holds where one of the input states is
the vacuum state and for several candidates of the other input state that includes the cases when the state has
the eigenvectors as the number states and either has only two non-zero eigenvalues or has arbitrary number of non-zero
eigenvalues but is a high entropy state. We also discuss the conditions, which if satisfied, would lead to an extension
of these results.
}

\section{Introduction}

The Entropy Photon Number Inequality (EPnI) was conjectured by Guha et. al. \cite{guha}.
EPnI has a classical analogue called Entropy power inequality which is stated as follows.
Let $X$ and $Y$ be independent random variables with densities and $h(X)$ be the differential entropy of $X$, then
\begin{equation}
\label{xy}
\myexp^{2h(X)}+\myexp^{2h(Y)} \leq \myexp^{2h(X+Y)}
\end{equation}
holds. It was first stated by Shannon in Ref. \cite{shannon} and the proof was given by Stam and Blachman \cite{stam,blachman}.

The EPnI has some important consequences in quantum information theory. In particular, if this conjecture
is true, then one would be able to establish the classical capacity of certain bosonic channels \cite{guha,Giovannetti}.
EPnI is shown to imply two minimum output entropy conjectures,  which would suffice to prove the capacity of
several other channels such as the thermal noise channel \cite{Giovannetti} and the  bosonic broadcast channel
\cite{guha-pra-2007,guha-isit-2008}.

The statement of the inequality is as follows. Let $a$ and $b$ be the photon annihilation operators and let the joint state of
the modes associated with $a$ and $b$ be the product state, i.e., $\rho_{AB} = \rho_A \otimes \rho_B$, where $\rho_A$
and $\rho_B$ are the density operators associated with the $a$ and $b$ modes respectively.
For the beam-splitter with inputs $a$ and $b$ and output $c$ with transmissivity $\eta$ and reflectivity $1-\eta$
respectively, the annihilation operator evolution is given by
\beq
c = \sqrt{\eta} a + \sqrt{1-\eta}b,
\enq
The  EPnI is now stated as
\begin{equation}
\label{bigepni}
g^{-1}\left[ S(\rho_C) \right] \geq \eta g^{-1}\left[ S(\rho_A) \right] + (1-\eta) g^{-1}\left[ S(\rho_B) \right],
\end{equation}
where
\beq
g(x) = (x+1) \log(x+1) - x \log(x)
\enq
is the von Neumann entropy of the thermal state with mean photon-number $x$, and
\linebreak $S(\rho) =$ $-\tr (\rho \log \rho )$ is the von Neumann entropy. 

In this paper, we prove the EPnI for the case of $\rho_B$ to be the vacuum state, $\rho_A$
having its eigenvectors as the number states and either having two nonzero eigenvalues or high von Neumann entropy
with arbitrary number of eigenvalues. There are other candidates as well for which some special cases
EPnI hold and these are mentioned  later.

\section{The beam-splitter transformation}

We obtain the output density matrix $\rho_C$ from the beam-splitter transformations.
The annihilation operators for the two outputs are
\begin{align}
\label{annh1}
c = \sqrt{\eta} a + \sqrt{1-\eta}b, \\
\label{annh2}
d = \myexp^{\iota\phi}(\sqrt{1- \eta} a - \sqrt{\eta} b),
\end{align}
where $[a,a^{\dagger}]=[b,b^{\dagger}]=[c,c^{\dagger}]=[d,d^{\dagger}] = \mathbb{I}$ and $[a,b]=[a,c]=[a,d]=0$ and so on.
We assume that the inputs density operators are diagonal in the number state basis and hence,
\begin{equation}
\rho_{AB} =  \sum_{i=0}^{\infty} \sum_{j=0}^{\infty} x_i y_j \ket{i}_A\ket{j}_B\bra{i}_A\bra{j}_B,
\end{equation}
where $x_i$ and $y_j$ are the $i$th and $j$th eigenvalues of $A$ and $B$ respectively,
$\ket{i}_A$ and $\ket{j}_B$ are the Fock number states for the systems $A$ and $B$ respectively.
Any state $\ket{i}_A\ket{j}_B$ can be written as (see Ref. \cite{gerry} for example)
\begin{equation}
\label{trans}
\ket{i}_A\ket{j}_B=  \frac{(a^\dagger)^i}{\sqrt{i!}}\frac{(b^\dagger)^j}{\sqrt{j!}} \ket{0}_A\ket{0}_B.
\end{equation}
From \eqref{annh1} and \eqref{annh2}, we get
$a^\dagger = \sqrt{\eta} c^\dagger  + \sqrt{1-\eta}\myexp^{\iota\phi}  d^\dagger$ and
$b^\dagger = \sqrt{1-\eta} c^\dagger  - \sqrt{\eta}\myexp^{\iota\phi}  d^\dagger$.
Using these with \eqref{trans}, we get the transformation
\begin{equation}
\ket{i}_A\ket{j}_B \xrightarrow{\ten{B.S.}} \frac{(\sqrt{\eta} c^\dagger  + \sqrt{1-\eta} \myexp^{\iota\phi} d^\dagger )^i}{\sqrt{i!}} \frac{(\sqrt{1-\eta} c^\dagger  - \sqrt{\eta}\myexp^{\iota\phi}  d^\dagger)^j}{\sqrt{j!}} \ket{0}_C \ket{0}_D,
\end{equation}
where B.S. indicates the action of the beam splitter.
Using the fact that the operators $c^\dagger$ and $d^\dagger$ commute and the binomial expansion, we get
\begin{align}
\ket{i}_A\ket{j}_B \xrightarrow{\ten{B.S.}} & \frac{1}{\sqrt{i!}\sqrt{j!}}  \sum_{k=0}^{i} \sum_{l=0}^{j}\myexp^{\iota (k+l) \phi} (-1)^{l} \binom{i}{k} 
\binom{j}{l} \eta^{\frac{i-k+l}{2}} (1-\eta)^{\frac{j-l+k}{2}}  \nonumber \\
& ~~~~ (c^\dagger)^{(i+j)-(k+l)} (d^\dagger)^{k+l} \ket{0}_C \ket{0}_D.
\end{align}
Incorporating the action of $c^\dagger$ and $d^\dagger$ on the vacuum states of $C$ and $D$, we get
\begin{align}
\ket{i}_A\ket{j}_B \xrightarrow{\ten{B.S.}} & \frac{1}{\sqrt{i!}\sqrt{j!}}  \sum_{k=0}^{i} \sum_{l=0}^{j} \myexp^{\iota (k+l) \phi} (-1)^{l} \binom{i}{k} \binom{j}{l} \eta^{\frac{i-k+l}{2}} (1-\eta)^{\frac{j-l+k}{2}} \nonumber \\
& \sqrt{[(i+j)-(k+l)]!(k+l)!} ~ \ket{(i+j)-(k+l)}_C ~ \ket{k+l}_D.
\end{align}
Hence, we arrive at the expression for $\rho_{CD}$ as
\begin{align}
\rho_{CD} = & \sum_{i=0}^{\infty} \sum_{j=0}^{\infty} x_i y_j \frac{1}{i!j!}  \sum_{k=0}^{i} \sum_{l=0}^{j}  \sum_{k'=0}^{i}
\sum_{l'=0}^{j}   \myexp^{\iota [(k+l)-(k'+l')] \phi} (-1)^{l+l'}   \binom{i}{k} \binom{j}{l}\binom{i}{k'} \binom{j}{l'} \nonumber\\ 
  & ~~ \eta^{i-\frac{k+k'}{2}+\frac{l+l'}{2}} (1-\eta)^{j-\frac{l+l'}{2}+\frac{k+k'}{2}} \nonumber \\
 & ~~ \sqrt{[(i+j)-(k+l)]!(k+l)!}  \sqrt{[(i+j)-(k'+l')]!(k'+l')!} \nonumber \\
& ~~ \ket{(i+j)-(k+l)}_C \ket{k+l}_D \bra{(i+j)-(k'+l')}_C \bra{k'+l'}_D.
\end{align}
Now, tracing out system D, we get
\begin{align}
\label{rhoc}
\rho_C = & \sum_{i=0}^{\infty} \sum_{j=0}^{\infty} x_i y_j \frac{1}{i!j!}  \sum_{k=0}^{i} \sum_{l=0}^{j}\sum_{k'=0}^{i}
\sum_{l'=0}^{j}   (-1)^{l+l'}   \binom{i}{k} \binom{j}{l}\binom{i}{k'} \binom{j}{l'} \nonumber \\
& ~~ \eta^{i-\frac{k+k'}{2}+\frac{l+l'}{2}} (1-\eta)^{j-\frac{l+l'}{2}+\frac{k+k'}{2}} \nonumber\\ 
& ~~ [(i+j)-(k+l)]!(k+l)!  ~ \ket{(i+j)-(k+l)} \bra{(i+j)-(k+l)} ~ \delta_{k+l,k'+l'}.
\end{align}

We now consider the special case when $\rho_B$ is a vacuum state.
Let the set of all probability vectors (with infinite length) be denoted by $\mathbb{P}$ and if $\pmb{x} \in \mathbb{P}$, then
$\sum_{i=0}^\infty x_i = 1$ and $x_i \geq 0$ $\forall$ $i \geq 0$.
Then \eqref{rhoc} reduces to
\beq
\rho_C = \sum_{i=0}^\infty z_i \ket{i}_C \bra{i}_C,
\enq
where
$\pmb{z} = M_\eta(\pmb{x}) \triangleq M(\eta,\pmb{x})$,
$M : [0,1] \times \mathbb{P} \to \mathbb{P}$ is a transformation given by
\begin{equation}
\label{meta}
z_i = \sum_{k=i}^\infty \binom{k}{i}\eta^i(1-\eta)^{k-i} x_k.
\end{equation}
Hence, \eqref{bigepni} reduces to
\beq
\label{epni}
g^{-1} \left\{ H[M_{\eta}(\pmb{x})] \right\} \geq \eta g^{-1} \left[ H(\pmb{x}) \right].
\enq
Note that this equation is expected to hold for all $\pmb{x} \in \mathbb{P}$ and $\eta \in [0,1]$. The inequality
is trivially true for $\eta = 0$ since $M_0(\pmb{x}) = [1,0,...]$ implying $H[M_0(\pmb{x})]=0$, and for $\eta = 1$ since
$M_1(\pmb{x}) = \pmb{x}$.

\section{$\rho_A$ is two-dimensional in the number state basis and $\rho_B$ is the vacuum state}

Let
\beq
H_b (p) \triangleq -p \log(p) - (1-p) \log (1-p)
\enq
to be the binary entropy of a two-point probability distribution $[p,1-p]$ with $0 \leq p \leq 1$.
Let the eigenvalues of $\rho_A$ given by the probability vector $\pmb{x}=[1-\alpha,\alpha,0,...]$.
Therefore, $ H(\pmb{x})=H_b(\alpha)$ and $ H[M_{\eta}(\pmb{x})]=H_b(\eta\alpha)$.
We now prove \eqref{epni} for the above case.

\bel
\label{lemma1}
For all $\eta \in [0,1]$ and $\alpha \in [0,1]$, we have
\beq
g^{-1} \left[ H_b (\eta \alpha) \right] \geq \eta g^{-1} \left[ H_b (\alpha) \right].
\enq
with equality if and only if $\eta \in \{0,1\}$ or $\alpha=0$.
\enl
\begin{proof}
One can see that  $g^{-1} \left[ H_b (\eta \alpha) \right] =\eta g^{-1} \left[ H_b (\alpha) \right]$
if $\eta \in \{0,1\}$ or $\alpha=0$.
In all other cases, we show that
\beq
g^{-1} \left[ H_b (\eta \alpha) \right] > \eta g^{-1} \left[ H_b (\alpha) \right].
\enq

Let $f(\beta) \triangleq g^{-1} \left[ H_b (\beta) \right]$.
The Lemma is equivalent to showing that $f(\beta)/\beta$ is a strictly decreasing function in $0<\beta \leq 1$.
Note that since
$g(\beta) = H_b (\beta) + 2 \left[ \log(2) - H_b \left( 1/2  + \beta/2 \right) \right]$
and $\log(2) >  H_b \left( 1/2  + \beta/2 \right)$ for all $\beta \in (0,1)$, hence $g(\beta) > H_b(\beta)$ for all $0 < \beta <1$.
Since $g$ is one-to-one and increasing, we have $g^{-1} \left[ H_b (\beta) \right] < \beta$ for all $0 < \beta <1$ or
$f(\beta) < \beta$ for all $0< \beta < 1$.

It is not difficult to see that
\beq
\frac{d}{d \beta} \frac{f( \beta)}{\beta} =
\frac{\log \left\{ (1-\beta)[1 + f(\beta)] \right\} }
{ \beta^2 \log \left[ \frac{1+f(\beta)}{f(\beta)} \right] }
\enq
and since, using $f(\beta) < \beta$ for all $0<\beta < 1$, it follows that
$(1-\beta)[1 + f(\beta)]<1$ for all $0<\beta<1$, hence,  $f(\beta)/\beta$ is a strictly decreasing function in $0<\beta \leq 1$.
\end{proof}

Recall that if the distribution of a random variable $X$ is Binomial,
denoted by $\ten{Bin}(L,\eta) \in \mathbb{P}$,
then $\ten{Bin}(L,\eta,k) \triangleq \Pr\{X=k\} = \binom{L}{k} \eta^k (1-\eta)^{L-k}$ if $k \in \{0,1,...,L\}$ and
is zero otherwise.

Let the two non-zero entries of the probability vector $\pmb{x}^{N,P}$ be at the $N$-th and $P$-th position, i.e.,
$x_N=1-\alpha,x_P=\alpha$ and let $\pmb{z}^{N,P}=M_{\eta}(\pmb{x}^{N,P})$.
\bel
For all $\eta \in [0,1]$, $\alpha \in [0,1]$ and $L \geq 1$, we have
\beq
g^{-1} \left[ H (\pmb{z}^{N,P}) \right] \geq \eta g^{-1} \left[ H(\pmb{x}^{N,P}) \right].
\enq
\enl
\begin{proof}
By Lemma \ref{lemma1}, we have
\beq
g^{-1} \left[ H_b (\eta \alpha) \right] \geq \eta g^{-1} \left[ H_b (\alpha) \right].
\enq
Note that $g$ is one-one and and strictly increasing, therefore $g^{-1}$ is also strictly increasing.
Therefore, it is enough to prove that
\beq
\label{dummys1}
H (\pmb{z}^{N,P}) \geq H (\pmb{z}^{0,1}).
\enq
as $H (\pmb{z}^{0,1})=H_b (\eta \alpha)$ and $H(\pmb{x}^{N,P}) = H_b(\alpha)$.
We first show that
\beq
\label{dummy1}
H (\pmb{z}^{0,P}) \geq H (\pmb{z}^{0,1}).
\enq
Note that
\begin{align}
H (\pmb{z}^{0,P}) = & f\left[ \alpha, (1-\eta)^P \right] + \alpha H \left[ \ten{Bin}(P,\eta) \right],
\end{align}
where
\beq
f(\alpha,x) = -\left[ (1-\alpha) + \alpha x \right] \log\left[ (1-\alpha) + \alpha x \right] - (1-x)  \alpha  \log(\alpha) + x\log(x) \alpha.
\enq
It is not difficult to show that $f(x)$ is a decreasing function of $x$. Note that $H \left[ \ten{Bin}(P,\eta) \right]$ increases
with $P$. Since $H (\pmb{x}^{0,P})$ is a sum of two functions each of which increases with $P$, \eqref{dummy1} follows.

Next, we show that for all $N, P \geq 0$, we have
\beq
\label{dummy2}
H (\pmb{z}^{N+1,P+1}) \geq H (\pmb{z}^{N,P}).
\enq
Note first that $\ten{Bin}(N+1,\eta) = (1-\eta) \ten{Bin}(N,\eta) + \eta \ten{Bin}_{+1}(N,\eta)$, where if $X$ has distribution
$\ten{Bin}_{+1}(N,\eta)$, then $\Pr\{X=k+1\} = \ten{Bin}(N,\eta,k)$ $\forall$ $k$. This implies that
\beq
\pmb{z}^{N+1,P+1} = (1-\eta) \pmb{z}^{N,P} + \eta \pmb{z}^{N,P}_{+1},
\enq
where we define $\pmb{z}^{N,P}_{+1}$ similarly. Using $H (\pmb{z}^{N,P}) = H (\pmb{z}^{N,P}_{+1})$,
it is not difficult to show that
\beq
H (\pmb{z}^{N+1,P+1}) = H (\pmb{z}^{N,P}) + (1 - \eta) D \left[ \pmb{z}^{N,P} || \pmb{z}^{N+1,P+1} \right]
+ \eta D \left[ \pmb{z}^{N,P}_{+1} || \pmb{z}^{N+1,P+1} \right],
\enq
where $D(\cdot||\cdot)$ is the relative entropy that is always non-negative and hence, \eqref{dummy2} follows.

Assume w.l.o.g. that $P > N$.
Applying \eqref{dummy1} repeatedly followed by \eqref{dummy2}, we get
\beq
H (\pmb{z}^{N,P}) \geq H (\pmb{z}^{0,P-N}) \geq H (\pmb{z}^{0,1}).
\enq
The result follows.
\end{proof}

\section{$\rho_A$ has number states as eigenvectors and $\rho_B$ is the vacuum state}

We have observed that the EPnI holds when $\rho_A$ has two non-zero eigenvalues with eigenvectors as the number
states and $\rho_B$ is a vacuum state. We
now consider the case  when $\rho_A$ has number states as the eigenvectors and could have arbitrary number of nonzero
eigenvalues and $\rho_B$ is the vacuum state.
We derive some necessary and sufficient conditions for this inequality to hold.

We first note that $M_{\eta} \left[ M_{\eta^\prime}(\pmb{x}) \right] =M_{\eta\eta^\prime}(\pmb{x})$
$\forall$ $\eta,\eta^\prime\in[0,1]$ and $\pmb{x} \in \mathbb{P}$. To prove this, let $\pmb{y} = M_{\eta^\prime}(\pmb{x})$, $\pmb{z} = M_{\eta}(\pmb{y})$
and note that
\begin{align}
z_i=& \sum_{k=i}^\infty \binom{k}{i} \eta^i (1-\eta)^{k-i} y_k \\
=&\sum_{k=i}^{j}\binom{k}{i}\eta^i(1-\eta)^{k-i}  \sum_{j=k}^{\infty} \binom{j}{k} {(\eta^\prime})^k{(1-\eta^\prime)}^{j-k}x_j\\
=&\sum_{j=i}^{\infty}\binom{j}{i}{(\eta\eta^\prime)}^i x_j
\sum_{k-i=0}^{j-i}\binom{j-i}{k-i}(\eta^\prime-\eta\eta^\prime)^{k-i}{(1-\eta^\prime)}^{j-k} \\
=&\sum_{j=i}^{\infty}M_{\eta\eta^\prime}x_j.
\end{align}

To simplify the notation, let us define
\begin{align}
H(\eta,\pmb{x}) & \triangleq H(M_{\eta} x) \\
h(\eta,\pmb{x}) & \triangleq g^{-1} \left[ H(\eta,\pmb{x})\right].
\end{align}
As $M_{1}$ is an identity transformation, we sometimes write $H(\pmb{x})$ for $H(1,\pmb{x})$ and $h(\pmb{x})$ for $h(1,\pmb{x})$.
Note that $h(1,\pmb{x})=g^{-1} \left[ H(\pmb{x}) \right]$ and therefore, \eqref{epni} can be rephrased as
\beq
\frac{h(\eta,\pmb{x})}{\eta}\geq  h(1,\pmb{x}).
\enq

It is not difficult to see that if \eqref{epni} holds, then
$h(\eta,\pmb{x})/\eta$ is a decreasing function in $\eta$.
To see this, let $\eta^\prime \leq \eta$ and $\delta = {\eta}^\prime/\eta$ where $0 \leq \delta \leq 1$. Then
\begin{align}
\frac{h(\eta^\prime,\pmb{x})}{\eta^\prime}& = 
\frac{h[\delta,M_{\eta}(\pmb{x})]}{\delta}\frac{1}{\eta}\\
&\geq \frac{h[1,M_{\eta}(\pmb{x})]}{\eta}\\
\label{dummy3}
&=\frac{h(\eta,\pmb{x})}{\eta}.
\end{align}
As $h(\eta,\pmb{x})/\eta$ is differentiable, we have
\beq
\frac{d}{d \eta} \frac{h(\eta,\pmb{x})}{\eta} = \eta \frac{d H(\eta,\pmb{x})}{d \eta}-H(\eta,\pmb{x}) + \log \left[ 1+h(\eta,\pmb{x}) \right].
\enq

\bel
\label{ineq}
Let $M_{\eta}: [0,1] \times \mathbb{P} \to \mathbb{P}$ be the transformation given by \eqref{meta}.
The following are equivalent:
\begin{align}
& (i) & ~ h(\eta, \pmb{x}) & \geq \eta h(1, \pmb{x}) & ~\forall ~ \pmb{x} \in \mathbb{P}, \forall ~ \eta \in (0,1], \\
& (ii) & ~ \frac{d}{d \eta} \frac{h(\eta,\pmb{x})}{\eta} & \leq 0 & ~\forall ~ \pmb{x}\in\mathbb{P}, \forall ~ \eta \in (0,1], \\
& (iii) & ~ \frac{d}{d \eta} \frac{h(\eta,\pmb{x})}{\eta} \Big\vert_{\eta=1} & \leq 0 & ~\forall ~ \pmb{x}\in\mathbb{P}.
\end{align}
\enl
\begin{proof}
It is clear from \eqref{dummy3} that $(i)$ and $(ii)$ are equivalent.
Furthermore, $(ii)$ implies $(iii)$ since $(iii)$ is a special case of $(ii)$. We prove that $(iii)$ implies $(ii)$.
Note that
\begin{align}
\label{dummy4}
\frac{d}{d \beta} \frac{h[\beta,M_\eta(\pmb{x})]}{\beta} \Big\vert_{\beta=1} & = \frac{d}{d \beta} \frac{h(\eta \beta,\pmb{x})}{\beta}
\Big \vert_{\beta=1} \\
\label{dummy42}
& = \eta^2 \frac{d}{d \theta} \frac{h(\theta,\pmb{x})}{\theta} \Big \vert_{\theta=\eta}.
\end{align}
Now $(iii)$ implies that
\begin{align}
\frac{d}{d \theta} \frac{h(\theta,\pmb{x})}{\theta} \Big \vert_{\theta=\eta} & \leq 0
\end{align}
and hence, $(ii)$ follows using \eqref{dummy42}.
\end{proof}

We now state EPnI in \eqref{epni} in the form of an  entropic inequality, i.e., an inequality involving Shannon entropy of discrete
probability distributions. By Lemma \ref{ineq}, \eqref{epni} is equivalent to
\begin{align}
\eta \frac{d H(\eta,\pmb{x})}{d \eta}-H(\eta,\pmb{x})+\log \left[ 1+h(\eta,\pmb{x}) \right] \leq 0.
\end{align}
The above can be expressed as
\beq
\label{temp1}
g \left[ \myexp^{H(\eta,\pmb{x})-\eta \frac{d H(\eta,\pmb{x})}{d \eta}}-1\right] \geq H(\eta,\pmb{x}).
\enq
Note that $g(1/\beta-1)=H_b(\beta)/\beta$ $\forall$ $\beta \in [0,1]$ and hence, \eqref{epni} is equivalent to showing that
\begin{align}
\label{dummy5}
H(\eta,\pmb{x})\leq \frac{H_b \left[ \myexp^{-H(\eta,\pmb{x})+\eta \frac{d H(\eta,\pmb{x})}{d \eta}} \right] }{\myexp^{-H(\eta,\pmb{x})+\eta
\frac{d H(\eta,\pmb{x})}{d \eta}}}.
\end{align}
For the two dimensional case with $\eta=1$, $\pmb{x} = [\alpha, 1-\alpha,0,...]$, $\alpha \in [0,1]$,
$H(\eta,\pmb{x})-\eta d H(\eta,\pmb{x})/d \eta=-\log(\alpha)$, $H(\pmb{x}) = H_b(\alpha)$, and substituting this in \eqref{dummy5},
we get
\beq
\label{temp5}
H_b(\alpha) \leq {H_b(\alpha) \over \alpha},
\enq
which is true. This gives a short proof of \eqref{epni} for this special case. Evaluating \eqref{dummy5} at $\eta = 1$ gives
an interesting expression that depends only on the distribution $\pmb{x}$. It is shown in \eqref{temp3} that
\beq
\Theta(\pmb{x}) \triangleq \frac{d H(\eta,\pmb{x})}{d \eta} \Big\vert_{\eta=1} = - \sum_{i=1}^{\infty} i x_i \log \left( \frac{x_i}{ x_{i-1}} \right),
\enq
and hence, \eqref{dummy5} reduces to
\beq
H(\pmb{x}) \leq \frac{H_b \left[ \myexp^{-H(\pmb{x})+\Theta(\pmb{x})} \right] }{\myexp^{-H(\pmb{x})+ \Theta(\pmb{x})}}.
\enq
The above inequality involves only entropies and another function $\Theta$ of the distribution but, to the
best of our knowledge, has never been studied before in the literature.

We now show that if \eqref{epni} is true, then it implies that
\begin{align}
\label{dummy6}
\eta \frac{d H(\eta,\pmb{x})}{d \eta} < 1, \\
\label{dummy7}
\eta \frac{d H(\eta,\pmb{x})}{d \eta} \leq H(\eta,\pmb{x}).
\end{align}
If \eqref{epni} holds, then using Lemma \ref{ineq}, we have
$H(\eta,\pmb{x})-\eta d H(\eta,\pmb{x})/d \eta \geq \log \left[ 1+h(\eta,\pmb{x}) \right]$.
As $\log \left[ 1+h(\eta,\pmb{x}) \right] \geq 0$, we have $H(\eta,\pmb{x})-\eta d H(\eta,\pmb{x})/d \eta \geq 0$, which
proves \eqref{dummy7}.

Using Lemma \ref{ineq} again, we have
$\eta d H(\eta,\pmb{x})/d \eta-H(\eta,\pmb{x})+\log \left[ 1+h(\eta,\pmb{x}) \right] \leq 0$.
It is enough to prove that $H(\eta,\pmb{x})-\log \left[ 1+h(\eta,\pmb{x}) \right] \leq 1$, i.e.,
\beq
1+g^{-1} \left[ H(\eta,\pmb{x}) \right] \geq \myexp^{H(\eta,\pmb{x})-1}.
\enq
We first consider the case when  $0\leq H(\eta,\pmb{x})\leq 1$. Then 
$\myexp^{H(\eta,\pmb{x})-1}\leq 1$. Therefore, $1+g^{-1} \left[H(\eta,\pmb{x})
\right] \geq \myexp^{H(\eta,\pmb{x})-1}$ and \eqref{dummy6} holds.

Now consider $H(\eta,\pmb{x}) \geq 1$.
Hence, it is enough  to prove that $1+g^{-1}(x)\geq \myexp^{x-1}$ $\forall$ $x\geq 1$, or,  $x+1\geq g(\myexp^x-1)$ $\forall$ $x \geq 0$.
Simplifying, we can show that this is equivalent to showing that $r(\myexp^{-x}) \geq 0$, where $r: [0,1] \to \mathbb{R}$ and
\beq
r(x) = x+(1-x)\log (1-x).
\enq
Note that  $r(0)=0$ and $dr(x)/dx=-\log (1-x)$ $\geq 0$ $\forall$ $x \in [0,1]$.
Therefore, $r(x)\geq 0$ $\forall$ $x \in [0,1]$ and \eqref{dummy6} follows.

\eqref{dummy6} and \eqref{dummy7} are the necessary conditions for \eqref{epni} to hold. We now show that they
both hold under general conditions.

\bel
\label{ineq2}
For all $\eta \in [0,1]$ and $\pmb{x} \in \mathbb{P}$, the following hold:
\begin{align}
\label{dummy8}
\eta \frac{d H(\eta,\pmb{x})}{d \eta} < 1, \\
\label{dummy9}
\eta \frac{d H(\eta,\pmb{x})}{d \eta} \leq H(\eta,\pmb{x})
\end{align}
with equality if and only if $M_\eta(\pmb{x}) = [1,0,...]$.
\enl
\begin{proof}
Let $\pmb{z}=M_{\eta}(\pmb{x})$ and using
\begin{align}
\eta \frac{d z_i}{d \eta} = iz_i-(i+1)z_{i+1},
\end{align}
we get
\begin{align}
-\eta \frac{d H(\eta,\pmb{x})}{d \eta}&=\eta\sum_{i=0}^{\infty} \left[ 1 + \log(z_i) \right] \frac{d z_i}{d \eta}\\
\label{temp3}
&=\sum_{i=1}^{\infty} i z_i \log \left( \frac{z_i}{ z_{i-1}} \right) \\
& \stackrel{a}{\geq} \sum_{i=1}^{\infty} i z_i \left( 1 - \frac{z_{i-1}}{z_i} \right) \\
\label{dummy10}
& = -1,
\end{align}
where in $a$, we have used the inequality that $\log(x) \geq 1- 1/x$ for all $x\geq 0$ with equality if and only if $x=1$.
If $\pmb{z}$ is such that $z_i \neq 0$ $\forall$ $i$, then it is impossible to have an equality in $a$ since equality would imply
$z_{i-1} = z_i$ $\forall$ $i$ and this would imply that $\sum_{i=0}^\infty z_i$ is unbounded.

If $\pmb{z}$ has a finite number of nonzero values say $\pmb{z} = [z_0,z_1,...,z_{L-1},0,...]$, then \eqref{dummy10} can
be further tightened as
\begin{align}
\label{temp2}
\eta \frac{d H(\eta,\pmb{x})}{d \eta} & \leq 1 - L z_{L-1}.
\end{align}
Hence, \eqref{dummy6} holds.

We now prove \eqref{dummy9} or equivalently
\beq
\Theta(\pmb{z}) = -\sum_{i=1}^{\infty} i z_i\log \left( \frac{z_i}{z_{i-1}} \right) \leq H(\pmb{z}).
\enq
Let us define a sequence of probability distributions $\{\pmb{z}^{(L)}\}$, $L=0,1,...$, where $\pmb{z}^{(L)}$ has length $L+1$ and
$\pmb{z}^{(L)} = [(1-z_L) \pmb{z}^{(L-1)}, z_L]$ and $\pmb{z}^{(0)} = [1]$. It is easy to see that the following recurrence
relations hold
\begin{align}
\label{dummys10}
\Theta(\pmb{z}^{(L)}) & = (1-z_L) \Theta(\pmb{z}^{(L-1)}) + L z_L \log \left( {1 - z_L \over z_L} z_{L-1} \right) \\
\label{dummy11}
H(\pmb{z}^{(L)}) & = (1-z_L) H(\pmb{z}^{(L-1)}) + H_b(z_L).
\end{align}
Define
\beq
\Xi(\pmb{z}^{(L)}) \triangleq \Theta(\pmb{z}^{(L)}) - H(\pmb{z}^{(L)}).
\enq
Using the recurrence relations in \eqref{dummys10} and \eqref{dummy11}, we get
\beq
\Xi(\pmb{z}^{(L)}) = (1-z_L) \Xi(\pmb{z}^{(L-1)}) + L z_L \log \left( {1 - z_L \over z_L} z_{L-1} \right) - H_b(z_L).
\enq
We now claim that
\beq
\label{dummy12}
\Xi(\pmb{z}^{(L)}) \leq L \log(1-z_L).
\enq
We prove this by induction. It is easy to check that $\Xi(\pmb{z}^{(1)}) = \log(1-z_1)$. Let \eqref{dummy12}
hold for $L-1$, $L > 1$. Then we have
\begin{align}
\Xi(\pmb{z}^{(L)}) & = (1-z_L) \Xi(\pmb{z}^{(L-1)}) + L z_L \log \left( {1 - z_L \over z_L} z_{L-1} \right) - H_b(z_L) \\
& \stackrel{a}{\leq} (L-1) (1-z_L) \log(1-z_{L-1}) + (L-1) z_L \log \left( z_{L-1} \right) +
L z_L \log \left( {1 - z_L \over z_L} \right) \nonumber \\
& ~~~ - H_b(z_L) \\
& \stackrel{b}{=} - (L-1) d(z_L,z_{L-1}) + L \log(1-z_L) \\
& \leq L \log(1-z_L),
\end{align}
where in $a$, we have used the induction hypothesis and the fact that $z_L \log(z_{L-1}) \leq 0$, in $b$,
\beq
d(x,y) = x \log \left( {x \over y} \right) + (1-x) \log \left( {1-x \over 1-y} \right)
\enq
is the relative entropy between $[x,1-x]$ and $[y,1-y]$ and is always nonnegative.
\eqref{dummy9} now follows from \eqref{dummy12} since $\log(1-z_L) \leq 0$. The equality condition follows straightforwardly.
\end{proof}

It is not difficult to see that the sufficient condition for \eqref{epni} to hold is that
$d H(\eta,\pmb{x})/d \eta \leq 0$. This condition is, of course, not true for many distributions such as a 
distribution whose sequence of entries are non-increasing.
Suppose $\pmb{z} = M_{\eta}(\pmb{x})$ has some zero entries in its interior, i.e., $z_i=0$ and $z_{i+1} \neq 0$
for some $i$. Then one can easily check that $d H(\eta,\pmb{x})/d \eta = -\infty$ and \eqref{epni} holds. It also follows from
\eqref{temp2} that if, for distributions with finite non-zero entries of the form $\pmb{z} = [z_0,z_1,...,z_{L-1},0,...]$
and $z_{L-1} \geq 1/L$,
then \eqref{epni} holds.

We now show that \eqref{epni} holds if $H(\pmb{x})$ is sufficiently large.
\bel
\label{lemma8}
For a given $\eta\in (0,1)$, $\pmb{x} \in \mathbb{P}$, \eqref{epni} holds if
$H(\pmb{x})$ is large enough.
\enl
\begin{proof}
Using \eqref{temp1}, we need to show that
\beq
g\left[ \myexp^{H(\eta,\pmb{x})-\eta d H(\eta,\pmb{x})/d \eta}-1 \right] \geq H(\eta,\pmb{x}).
\enq
We have
\begin{align}
g\left[ \myexp^{H(\eta,\pmb{x})-\eta d H(\eta,\pmb{x})/d \eta}-1 \right] & \stackrel{a}{>} H(\eta,\pmb{x}) + \delta - \myexp^{-
H(\eta,\pmb{x}) + \eta d H(\eta,\pmb{x})/d \eta} \\
& \stackrel{b}{>} H(\eta,\pmb{x}) + \delta - \myexp^{- H(\eta,\pmb{x}) + 1} \\
& \geq H(\eta,\pmb{x}),
\end{align}
where in $a$, we use the inequality that $g(\myexp^x-1)\geq x+1-\myexp^{-x}$ and we use Lemma \ref{ineq2}
to get $\eta d H(\eta,\pmb{x})/d \eta < 1 - \delta$ for some $\delta > 0$, in $b$, we use $\eta d H(\eta,\pmb{x})/d \eta < 1$ and
the last inequality would hold if $H(\eta,\pmb{x}) \geq 1 - \log(\delta)$ or if $H(\eta,\pmb{x})$ is large enough.

We now show that if $H(\pmb{x})$ is large, then so is $H(\eta,\pmb{x})$ for $\eta \in (0,1)$. Define
\beq
q(\eta,\pmb{x}) \triangleq { H(\eta,\pmb{x}) \over \eta }.
\enq
Differentiating w.r.t. $\eta$, we get using \eqref{dummy9},
\begin{align}
{d q(\eta,\pmb{x}) \over d \eta} & = {1 \over \eta^2} \left[ \eta \frac{d H(\eta,\pmb{x})}{d \eta} - H(\eta,\pmb{x}) \right] \\
& \leq 0.
\end{align}
Hence, $q(\eta,\pmb{x})$ is a decreasing function of $\eta$ and $H(\eta,\pmb{x}) \geq \eta H(\pmb{x})$. Similarly, using \eqref{dummy8},
we get
\begin{align}
\int_{\eta}^1 d H(\beta,\pmb{x}) < \int_\eta^1 {d \beta \over \beta} \\
H(\eta,\pmb{x}) > H(\pmb{x}) + \log(\eta).
\end{align}
Hence,
\beq
H(\eta,\pmb{x}) \geq \max \left\{ \eta H(\pmb{x}), H(\pmb{x}) + \log(\eta) \right\}.
\enq
This shows that if $H(\pmb{x})$ is large, then so is $H(\eta,\pmb{x})$ and hence, \eqref{epni} would hold for any
$\eta \in (0,1]$ for large $H(\pmb{x})$.
\end{proof}

\section{Discussion}

It is, of course, of great interest to see if these results could be generalized for the cases where $\rho_A$ and $\rho_B$ do
not have the special structure such as the eigenvectors being the number states etc. It would seem that our results may
extend over to cover some of these cases if the following is established. Suppose there exists an $\pmb{x} \in \mathbb{P}$
such that
\beq
\frac{d}{d \eta} \frac{h[\eta,M_\beta(\pmb{x})]}{\eta} \Big\vert_{\eta=1}  < 0  ~\forall ~ \beta  \in  (0,1].
\enq
Then, it follows from \eqref{dummy42} that
\beq
\frac{d}{d \eta} \frac{h(\eta,\pmb{x})}{\eta} \Big\vert_{\eta=\beta}  < 0  ~\forall ~ \beta \in (0,1].
\enq
This would then imply that $h(\beta,\pmb{x})$ is a strictly decreasing function of $\beta  \in  (0,1]$ and hence,
\eqref{epni} holds with strict inequality.

An example of such a $\pmb{x}$ 
is $\pmb{x} = [\alpha, 1-\alpha,0,...]$, $\alpha \neq 1$, and $M_\beta(\pmb{x}) = [1 - (1-\alpha)\beta, (1-\alpha) \beta,0,...]$,
and \eqref{epni} is strict using \eqref{temp5}.

For finite $n$ and any state $\sigma$ defined on the number states as
\beq
\sigma = \sum_{i,j=0}^n \xi_{i,j} \ket{i} \bra{j},
\enq
we define a function
\beq
f(n,\sigma) = \sum_{i,j=0}^n \xi_{i,j} \ket{e_i^n} \bra{e_j^n},
\enq
where $\{\ket{e_i^n}\}$ is the standard basis for the Hilbert space of dimension $n+1$, i.e.,
$\bra{e_i} = [\stackrel{i}{\overbrace{{0,...,0}}},1,\stackrel{n-i}{\overbrace{0,...,0}}]$, $i=0,1,...,n$.
It follows that $S(\sigma) = S[f(n,\sigma)]$.

Now consider the input states such that
\begin{align}
\rho_A & = \sum_{i,j=0}^{n_A} \lambda_{i,j} \ket{i}_A \bra{j}_A, \\
\rho_B & = \sum_{i,j=0}^{n_B} \gamma_{i,j} \ket{i}_B \bra{j}_B, \\
\hat{\rho}_A & = \alpha \ket{0}_A \bra{0}_A + (1 - \alpha) \ket{1}_A \bra{1}_A \\
\hat{\rho}_B & = \ket{0}_B \bra{0}_B,
\end{align}
where $n_A,n_B$ are finite and
$|| f(n_A,\rho_A) - f(n_A,\hat{\rho}_A) ||_\mathrm{tr} < \delta$ and
$|| f(n_B,\rho_B) - f(n_B,\hat{\rho}_B) ||_\mathrm{tr} < \delta$.

It is not difficult to see that under the action of $f$, the output $\rho_C$ of beam splitter with $\rho_A$ and $\rho_B$ as inputs
is close to the output $\hat{\rho}_C$ with $\hat{\rho}_A$ and $\hat{\rho}_B$ as inputs, i.e.,
$ || f(n_A+n_B,\rho_C) - f(n_A+n_B,\hat{\rho}_C) ||_\mathrm{tr} < \epsilon$, where we could
make $\epsilon$ as small as possible by choosing $\delta$ small.
Using Fannes' inequality \cite{fannes-1973,nielsen}, this would result in a small deviation in the von Neumann entropies of
$\rho_A$, $\rho_B$ and $\rho_C$ as compared to $\hat{\rho}_A$, $\hat{\rho}_B$ and $\hat{\rho}_C$ respectively
that can be absorbed while still preserving the inequality since the inequality is strict.

\bibliographystyle{IEEEtran}
\bibliography{epni}

\end{document}